Title: Migration of photogenerated charge carriers in silver halides: small polaron transport
Author: Mladen Georgiev (Institute of Solid State Physics, Bulgarian Academy of Science
  1784 Sofia, Bulgaria)
Comments: 13 pages including 4 figures and 1 table, all pdf format
Subj-class: physics

In the late 60s of past century, then a young physicist, I joined an ambitious research project aimed at establishing the mechanism of photodecomposition of ionic salts. We measured the mobilities and lifetimes of photoelectrons and photoholes of silver halides in order to get an early-stage information on the process. The last pair of quantities to study was that of the respective diffusion coefficients for which we obtained results, highly controversial at that time. Now we regard the diffusion data to have suggested the formation of itinerant small polarons as the primary photocarriers in silver halides.

1. Introduction

Investigating the migration of photoexcited carriers in silver halides, one can arrive at definite conclusions on the mechanisms of the subsequent dissociation processes. Generally, once in the conductive bands, photoelectrons and photoholes may be expected to couple strongly to the environment, and indeed, this will ultimately trigger the dissociation channels. The direct result of environmental coupling will be the formation of electron and hole polarons, that is, free or bound carriers coupled to the lattice vibrations. In a medium with Jahn-Teller (JT) interactions, the migrating carriers might most likely create and couple self-consistently to the JT distortions resulting in vibronic or JT polarons. The polarons would be seen to survive decomposition trends as they traverse the crystal. Part of them would also couple to migrating interstitial silver ions to form silver specks inciting the photodecomposition channel of the salt. A summary of the discussion to follow has already been presented in an earlier arXiv paper [1].

1.1. A brief retrospective

In 1960 Jordan Malinowski (Sofia) and Peter Süptitz (Berlin) have jointly masterminded ingenious methods for the detection of photoelectrons and photoholes generated by strongly absorbed light at one (basal) illuminated surface and brought to the opposite (upper) detecting surface by electric fields or concentration diffusion. The detection method for photoelectrons consisted in pre-activating the surface by a chemical sensitizer to produce a latent image speck upon the arrival of an electron. For photoholes, the activation was by the preliminary deposition of a silver layer to be destroyed by emerging holes, the affected areas being easily revealed by "chemical development". Both wedge-shaped and tablet-like samples were used as strongly-absorbed ultraviolet light pulses were synchronized with field pulses for electric-drift measurement and no pulses for diffusion [2-6].

Normally rectangular light slits were carved on the illuminated surface which projected practically unchanged on the detection surface in electric drift experiments. Unlike it, diffusing holes produced elliptic-shaped silver bleachings, while diffusing electrons produced elliptic-shaped silver images intensified through development. (Elliptic-shaped was the cross-section of the wedge plane with a cylindrical diffusion front.) Sometimes, the internal parts of the latter images were bleached by co-migrating holes, though never the hole bleaching was complete, as expected for phase-segregated species. For determining the diffusion constants, we measured the elliptic semiaxes as functions of the time of exposure $t$ which following a curved part turned rectilinear $\propto \sqrt{t}$ at larger exposures. The slopes of the rectilinear portions were predicted by the theory to be $\propto D^{1/2}$ whereby the diffusion constant was easily deduced.

Malinowski&Süptitz's (MS) detection technique has also been applied with alternating degree of success to other salts too, including ionic and covalent solids. By extending the method to "physical development", the diversity of activated materials for photohole studies increased.

### 1.2. Migration of primary photocarriers

In a series of experiments, we measured the migration characteristics, such as the drift mobility and the apparent diffusion coefficient of species photogenerated at one of the surfaces of wedge-shaped single crystal specimens of AgBr. Using photographic detection methods we observed the appearance of what we tentatively called "photoelectrons" and "photoholes" at the opposite face of the crystal when driven by an electric force or by concentration diffusion. For "photohole" detection we followed their bleaching of thin (mono atomic) silver layers observed by means of photographic development [2-4]. For detecting the "photoelectrons" we deposited surface layers of an appropriate sensitizer and applied photographic development thereafter [5,6]. The method employed was measuring the penetration depth of the species. Much to our surprise, the diffusion constants so determined were largely inferior to estimates based on the drift mobility by several orders of magnitude, six for the "photoholes" and seven for the "photoelectrons", as seen in Figures 1 and 2 [2,5].

The obtained drift and diffusion data and their discrepancies indicated clearly that while the observed drifting carriers were the same as the photoelectrons and the photoholes generated by the illumination, the species involved in the slower diffusion motion were not. Most likely the diffusing species were kind of small polarons or bipolarons formed as the migrating electrons and holes coupled strongly to their deformed environment. We put these polaronic "photoelectrons" and "photoholes" in quotation marks. The small polaron formation merely provides a mechanism for a slow adiabatic carrier transport across the sample. Inasmuch as the formation of a small polaron takes time, the polarons do not suffice to form for too short field pulses and traversal times. This explains why the carriers in a drifting experiment have been exactly as generated by the light pulse. Unlike them, the diffusing carriers have certainly had enough time to form small polarons, except for a small quantity at the beginning of illumination.

From the measured discrepancy for the hole small polarons, $10^{-7} = \exp(-2E_{LR}/\hbar\omega)$ we get for the lattice relaxation energy $E_{LR} = 8\hbar\omega$ ($\hbar = h/2\pi$) which signifies considerable electron-phonon coupling strength. As a matter of fact, taking $\hbar\omega = 25$ meV we obtain $E_{LR} = 0.2$ eV, quite acceptable from the viewpoint of the small polaron theory. Similarly, we find $E_{LR} = 9.2\hbar\omega$ for the electron small polaron which implies an even higher lattice relaxation energy of $E_{LR} = 0.25$ eV. Given the above estimates for the lattice relaxation energies, it is not surprising that small polarons form so effectively in silver halides.

Now, inasmuch as the hole is known to self trap stably with cubic symmetry in AgCl but not in AgBr, the small polaron formation in AgBr requires further consideration. If the hole does not self trap stably, then it would form transient small polarons in AgBr. Once released to its conductive band, the hole would migrate over a distance, then self-trap temporarily again at another cubic site and so on. The sequence of self-trapping & untrapping steps is remindful of the multiple trapping known from semiconductor physics and tends to reduce the apparent mobility or diffusion coefficient outright, since the carrier remains immobile for a certain period ot its lifetime [7].

It can be argued further that the "photoelectrons" as small polarons otherwise centered at cubic sites may incorporate interstitial silver ions in their immediate distorted environment ($C_{4v}$ symmetry) and thereby stimulate the formation of transient silver atoms $Ag_1$ or diatoms $Ag_2$. The interstitial ions are expected to provide but transient electron traps. The "photohole" small polarons are chemically reactive as bromine atoms $Br_1$ at cubic sites or bromine molecules $Br_2$ if they occur in pairs.

Another essential conclusion to be drawn from our experiments is that the photochemical products associated with the electrons and the holes are phase segregated, as observed in high-$T_C$ superconducting materials lately [8]. Accordingly, phase segregation appears to be the basic prerequisite for the halide to decompose photochemically by preventing or lessening the effect of electron-hole recombination.

In a systematic study we could measure the barrier surmounted by the diffusing "photohole", as seen in Figure 2 [3], even though we failed technically to do so for the "photoelectron". The hole barrier can be made use of while addressing the nature of the diffusing "photoholes". Unfortunately no spectroscopic measurements on the primary electron or hole polarons were made to verify their exact nature. Nevertheless, our migration experiments lent support to the fine-structure hypothesis for the silver halides. Indeed, only small-size silver or halogen particles could traverse the bulk of a macro-specimen to appear on the side opposite to the illuminated face, as observed. These small particles should be observable in a time resolved spectroscopic experiment.

It would be fair to stress once again that alternative to the itinerant though stable small polaron is the multiple trapping model. In the latter, an immobile entity forms temporarily at a site which reduces the displacement distance by holding the migrating electrons or holes trapped for some time before releasing them again for a subsequent migration step. This multiple trapped entity is indistinguishable from the small-polaron as far as the temperature dependence of the diffusion coefficient is concerned although otherwise there is a clear distinction between the two in that the latter is slowly itinerant all the time, while the trapped entity is not. Apart from this, their external appearance is similar.

Microscopically, a "photoelectron" could be involved in a multiple trapping sequence by some more or less identified traps having no obvious link with the silver specks. In a particular case, however, the trapped entity could well be an unstable silver atom forming temporarily at the trapping site, e.g. by combining photoelectrons with interstitial silver ions, as in e' + $Ag_n$ + $Ag_i$ ↔ $Ag_{n+1}$. Alternatively, the small uni- or bi-polaron complexes may also have an obvious relationship with the silver specks or bromine molecules for this is what comes out to the surface for detection.

Table I

Diffusion coefficients of photoelectrons and photoholes from direct and drift mobility measurements in AgBr single crystals at room temperature

| Electrons | | Holes | |
|---|---|---|---|
| From drift mobility | From diffusion data | From drift mobility | From diffusion data |
| $D_e = \mu_e(k_B T/e)$ $\mu_e = 70$ cm$^2$V/s | $D_n =$ $1.5 \times 10^{-8}$ cm$^2$/s | $D_h = \mu_h(k_B T/e)$ $\mu_h = 1$ cm$^2$/s | $D_p =$ $2.6 \times 10^{-7}$ cm$^2$/s |
| $D_e =$ 1.8 cm$^2$/s | $D_n/D_e = 10^{-8}$ | $D_h =$ 0.026 cm$^2$/s | $D_p/D_h = 10^{-5}$ |
| | | | |

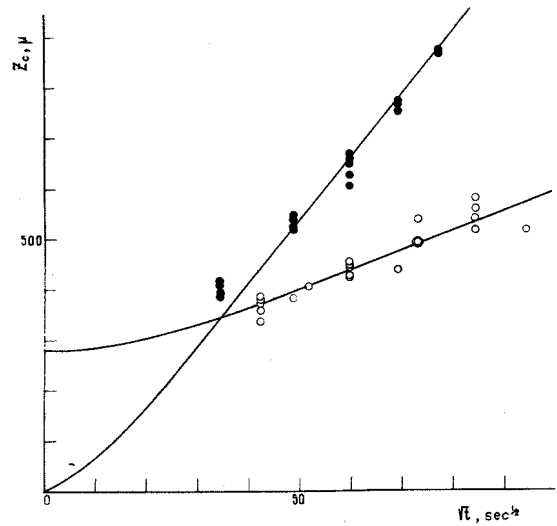

Figure 1: The diffusion paths of "photoelectrons" (hollow circles) and "photoholes" (filled circles) in silver bromide at room temperature. The linear parts at longer exposure times $t$ are proportional to the respective square root diffusion coefficients. The numerical estimates were $D_n = 1.5 \times 10^{-8}$ cm$^2$/s for the electrons and $D_p = 2.6 \times 10^{-7}$ cm$^2$/s for the holes. The diffusing entities were regarded as primary photocarriers in the formation of silver specks. The finite value of $z_C$ at $t = 0$ may be due to an initial process, e.g. self trapping, running faster for the photoelectrons than it does for the holes. (From Ref. [5].)

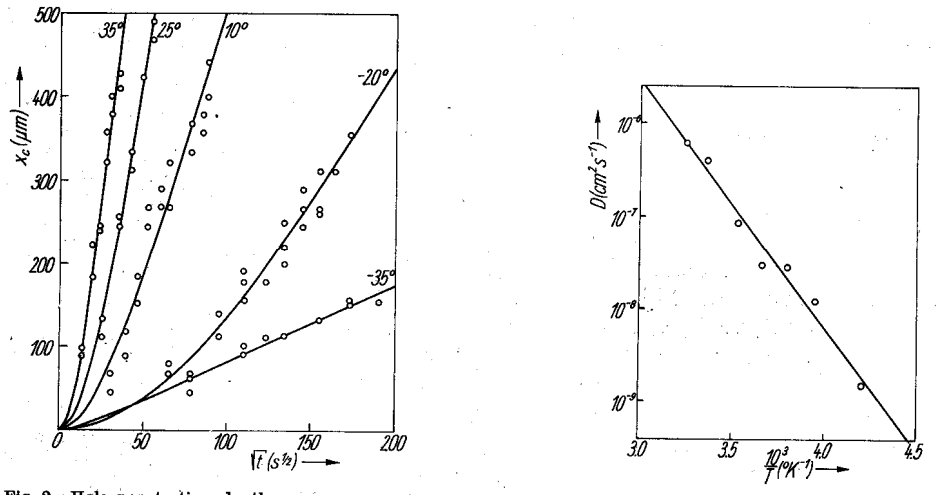

Figure 2: Left- diffusion paths (depth $x_C$ vs. square root of migration time $\sqrt{t}$ ) for photoholes in silver bromide between ±35°C. The diffusion constants were determined from the slopes of the rectilinear portions proportional to $\sqrt{D}$. Right- the derived Arrhenius temperature line following the equation $D(T) = \exp(5.44 - 0.525\ eV/k_BT)$, $cm^2$/s. From Ref. [3].

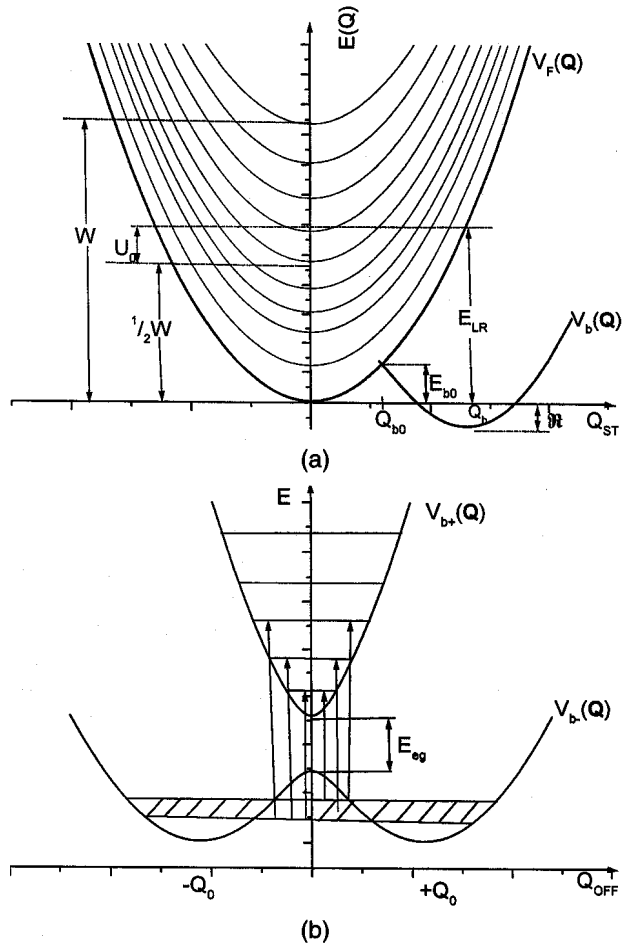

Figure 3: (a)- The sea of free-band states and a split self-trapped state. The carrier self traps stably if its lateral well falls below the free band minimum and the carrier is itinerant otherwise. (b)- The 1D diagram of the double-branch adiabatic potential. Note the crossover gap energy at Q = 0 (avoided crossing) which makes the interwell transitions across the barrier possible at all. For Jahn-Teller polarons, the crossover energy gap hosts states of the component bands, while the subbarrier polaron band (hatched) is composed of mixed states originating from the gap states squeezed by the phonon coupling (Holstein's reduction). From Ref. [7].

### 1.3. Temperature dependence of diffusion constant [10]

#### 1.3.1. Halogen emission from the dark surface

Nearly anyone having experimented with ionic crystals is aware of the typical halogen smell that appears when they are subjected to ionizing radiation. Among the common examples are gamma-rayed alkali halides and violet-light illuminated silver halides. Quite a few have asked the question just what the halogen transport mechanism to the surface is. Quantitatively photoinduced halogen evolution rates from silver halides have been measured by Luckey in 1962 to derive bulk diffusion coefficients, desorption from the surface being assumed a process much faster than bulk migration. It has later been realized that the difusion constants measured e.g. in AgBr (the order of $10^{-6}$ cm$^2$ s$^{-1}$) fall short by at least four orders of magnitude to agree with ones calculated from Hall mobility data using the Einstein relation (the order of $10^{-2}$ cm$^2$ s$^{-1}$) [5]. Comparable diffusion data have also been obtained on AgBr by photoexcitation experiments suggesting that the discrepancy might even be deeper [6]. Surprisingly, very low diffusion constants misfitting the mobility by some eight orders of magnitude have also been reported for photoexcited electrons in AgBr samples [5].

Although the misfit has at the time been attributed to muliple trapping reducing the drift mobility to low values, the traps have never been thoroughly identified (for a later review see Kanzaki in 1983) [11]. Another suggestion that the complex of a hole bound to some lattice defect (e.g. neutral cation vacancy) might be the slowly mobile entity has virtually remained unnoticed, since the $V_F$ center has lacked an optical evidence for the silver halides.

At nearly the same time, however, people have been aware that a hole does self trap in AgCl, even though it does not in AgBr. The STH in AgCl is actually a JT polaron by virtue of the vibronic coupling of Ag$^+$ - ion states. Bipolarons, just introduced during the mid sixties, have not been considered a real possibility. Nevertheless, diffusion and mobility data may also be reconciled in terms of small bipolarons on the halogen sublattice: Indeed, from negative-U theory we easily get

$$D_h / D_b = t_h / t_b =$$

$$(U_b / 2 t_h)(t_h / t_p)^2 = (U_b / 2 t_h) \exp(4E_{JT}/\hbar\omega); \qquad (1)$$

for the tunneling range which can eventually produce free-hole to hole-bipolaron diffusion-constant ratios largely exceeding unity. However, any consistent bipolaron theory of halogen migration in silver halides should take into account the particular range of experimental measurements which is thermally-activated rather than tunneling [3].

In the above-mentioned experiments, free carriers have been generated through strongly absorbed photoexcitation across an indirect bandgap, and then diffused to the opposite surface, where the evolving halogen has been detected [2]. The reason why photoelectrons and photoholes have apparently migrated separately, at least in measurable numbers, rather than recombining outright has always been a puzzle. At the same time, bandgap excitation of alkali halides results in the emission of halogen (and also alkali) atoms (see Berry 1973 for a later review)[12]. The surface evolution rate being proportional to the intensity of irradiation, a single excitation has been assumed responsible. Model proposals have included free excitons, self-trapped excitons, and self-trapped holes as species whose migration to the surface incites the desorption of atoms. However there seems to be no direct experimental proof for any of the related mechanisms.

### 1.3.2. Barrier controlled transition probabilities

Assuming coupling to a phonon mode and applying the *universal occurrence probability formula* for the rate of a barrier-controlled process, the diffusion coefficient of an α-species is

$$D_\alpha(T) = \Lambda_\alpha^2 (1/\tau_\alpha) = \Lambda_\alpha^2 \Sigma_n W_n(E_n) A_n(E_n) (1/Z_O) \exp(-E_n/k_BT) \qquad (2)$$

where $(1/Z_O) \exp(-E_n/k_BT)$ derives from Boltzmann's statistics, $W_n(E_n)$ is the probability of traversing the migrational barrier $E_B$ for diffusion. $A_n(E_n)$ is the adiabaticity factor, $E_n$ is the quantum energy spectrum coupled to the migrating species [10]. For phonon coupling $E_n = (n + ½)h\nu$, as the coupling results in displaced oscillators forming a barrier in-between, as in Figure 3.

In $D_\alpha$ some of the energy levels are underbarrier ($E_n < E_B$), while others are overbarrier ($E_n > E_B$). For underbarrier ($E_n \ll E_B$) [13],

$$W_n = \pi\{[F_{nm}(\xi_0,\xi_C)]^2 / 2^{n+m} n!m!\}\exp(-(n-m)^2 h\nu/E_R)\exp(-(E_R/h\nu)), \quad A_n = 2\pi\gamma^{2\gamma-1}\exp(-2\gamma)/\Gamma(\gamma)^2$$

with

$$F_{nm}(\xi_0,\xi_C) = \xi_{m0}H_n(\xi_C)H_m(\xi_C-\xi_{m0}) - 2nH_{n-1}(\xi_C)H_m(\xi_C-\xi_{m0}) + 2mH_n(\xi_C)H_{m-1}(\xi_C-\xi_{m0}), \qquad (3)$$

for overbarrier ($E_n \gg E_B$),

$$W_n \sim 1, \quad A_n = 2[1 - \exp(-2\pi\gamma)] / [[2 - \exp(-2\pi\gamma)]$$

with

$$\gamma_n(E_n) = (V_{12}/2h\nu) \, 1 / \sqrt{[E_R|E_n - E_C|]} \qquad (4)$$

standing for Landau-Zener's parameter. On the other hand. $1/Z_O = 2\sinh(½ \, h\nu/k_BT)$ for a single harmonic mode. The diffusion step occurs across the barrier through quantum tunneling or classical jump.

### 1.3.3. Zero-point rate

The energy gap splitting $V_{12} = ½E_{band} \ll E_B$ for $E_n \ll E_B$. Other important parameters are: barrier $E_B = E_{LR}(1 - \eta)^2$, $E_{LR}$ - lattice relaxation energy, $E_C = E_B + V_{12}$ crossover energy, zero-point reaction heat $Q = (m-n)h\nu$, lattice relaxation energy $E_R = 2Kq^2$ at $Q = 0$, etc. Here and above $\eta = E_{band}/4E_{JT}$ is the reduced gap energy of the polaron ensemble.

It is remarkable that the above formulae for $Z_O$ allow for the existence of a zero point diffusion constant given by:

$$D(0) = \Lambda^2 \Re(0) = \Lambda^2 \nu(E_R / h\nu) \exp(-E_R / h\nu) \qquad (5)$$

while the overall temperature dependence is described by

$$D(T) = \Lambda^2 \Re(T) = \Lambda^2 \Sigma_n W_n(E_n) A_n(E_n) (1/Z_O) \exp(-E_n/k_B T) \tag{6}$$

where $\Lambda$ is a diffusive area constant. The temperature dependence of the diffusion constant $D(T) = \Lambda^2 / \tau = \Lambda^2 \Re(T)$, where $\Re(T) = 1 / \tau$ is the diffusion rate, is shown in Figure 4. The calculations are made based on the estimates in Section 2.1 setting $E_R \sim 0.775$ eV at $E_{band} = 1.325$ eV, the bromine carrier bandwidth. The remaining parameters used are $E_B = 0.525$ eV from the Arrhenius slope, and tentatively $\eta \sim E_{band}/4E_B = 0.5$ or $0.05$.

Indeed, just to see where we are, from $\exp(5.44 - 0.525 \text{ eV}/k_B T_0) = \Lambda^2\Re(0)$ we get anyway: $T_0 \sim 0.525$ eV $/\{5.44 - \ln[\Lambda^2\Re(0)]\}$ $k_B$ which solves to $T_0 \sim 110$ K (typical for many diffusing agents) at $L^2\Re(0) = \exp(-50) \sim 10^{-22}$ cm$^2$/s. On the other hand, the theoretical formula gives for $\Lambda^2\Re(0) = 10^{-22}$ s$^{-1}$ and $\nu = 3.8 \times 10^{13}$ s$^{-1}$, $\Lambda^2 = 10^{-4}$ cm$^2$, $(E_R / h\nu) \exp(-E_R / h\nu) = \frac{1}{4}10^{-27}$ $(E_R / h\nu)$. For $\Re(0) = 10^{-18}$ cm$^2$.s$^{-1}$, $\nu = 4 \times 10^{13}$ s$^{-1}$, we get $\exp(-E_R / h\nu) \sim 10^{-31}$, $E_R \sim 0.775$ eV.

The lattice reorganization energy, $E_R = 4E_C = 4E_B + 4 \frac{1}{2}E_{band} \sim 2.1 - 2E_{band} = 0.775$ eV wherefrom $2E_{band} \sim 1.325$ eV, $E_{band} = 0.66$ eV. $E_{band}$ enters as the polaron bandwidth (halfwidth of 0.33 eV). The splitting of the electronic states at crossover is provided by the electronic bandgap $E_{band}$ – this turns the configuration avoided-crossing. The middle polaron band energy relative to the middle carrier bandgap is $E_P = -\frac{1}{2} E_{band} - E_{JT} + \frac{1}{2} h\nu$. The narrowing of the small polaron band at $E_P$ is given by $\Delta E_P = E_{band} \exp(-2E_{JT}/h\nu)$.

## 2.1. Carrier bands

We remind that the carrier band is composed of unihole or dihole states. Accordingly, the carrier is either an itinerant $Br_2$ molecule on an anion lattice site, or more specifically, an itinerant $Br_2^-$ molecular ion similar to the species in alkali halides. Both events take place on the negative-ion sublattice. Hole events on the cation sublattice are perhaps more intriguing to decipher because of the $Ag^{2+}$ hole self trapping which is not that usual. That hole carried by a JT ion, it will produce bound self-trapped polarons if in AgCl or itinerant self-trapped polarons if in AgBr, as in Figure 3 (a). By coupling to an even parity symmetry-breaking mode, such as $E_g$ and $T_{2g}$ from the symmetry group $O_h$, the phonon mixing lifts the degeneracy of the electronic system.

The electronic counterpart moves across the $Ag^+$-ion sublattice as well. We may argue that the photoelectron traps to form unstable $Ag^0$ atoms, not that unstable perhaps, since the reduction of the diffusing constant is $10^{-8}$ relative to the untrapped decoupled electron. A species staying trapped that long should not be called unstable anyway. The electronic configuration of Ag (#47 at.w.107.880) is [2,8,18,18,1] and that of Br (#35 at.w.79.916) is [2,8,18,7]. Upon forming a chemical bond, the unpaired electron of $Ag_{Ag}^\times$ is shared with $Br_{Br}^\times$ adding a partial covalent mixture to the basic ionic bond to build up the Kr [2,8,18,8] configuration. The electronic configurations are: $Ag^+(Ag_{Ag}^\times)[2,8,18,18]$, $Ag^{++}(Ag_{Ag}^\bullet)[2,8,18,17]$, $Br^-(Br_{Br}^\times)[2,8,18,8]$.

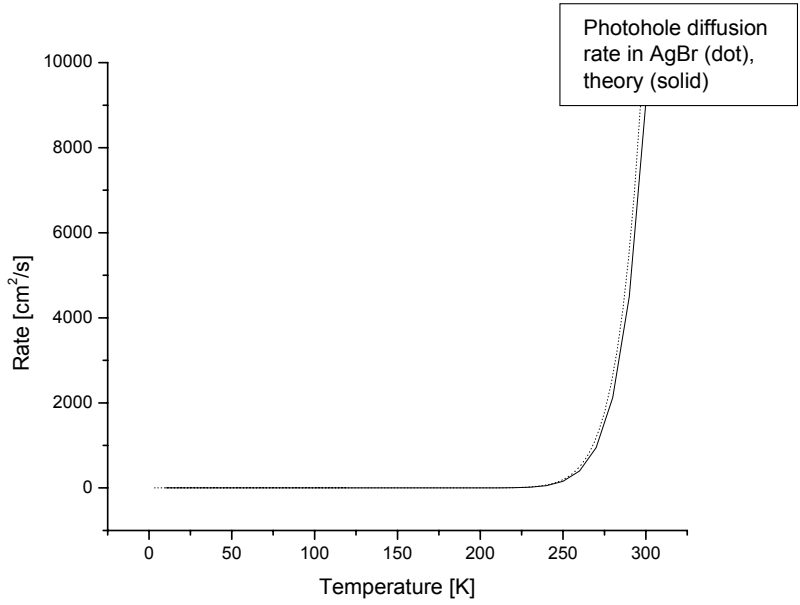

Figure 4: Photohole diffusion experimental in AgBr (dots) as compared with the quantal rate theory (solid). To reconcile rate with diffusion we used a factor of $\Lambda^2 = 10^{-10.5}$ cm$^2$. The remaining parameters of the theoretical fit are: $E_B = 0.525$ eV, $E_R = 10.5$ eV, $E_C = 2.625$ eV, $E_{band} = 4.2$ eV, $h\nu = 0.025$ eV, $\eta = E_{band}/4E_{JT} = 0.5$ which gives $E_{JT} = \tfrac{1}{2} E_{band} = 2.1$ eV. We see a broad low-temperature diffusion branch extending to ~ 100 K, typical for many diffusing agents.

## 2.2. Holstein polarons vs. Jahn-Teller polarons

For a Holstein polaron [14], the polaron band middle energy relative to the middle free-carrier bandgap is:

$$E_{P0} = -\tfrac{1}{2} E_{band0} - E_{JT0} + \tfrac{1}{2} h\nu, \qquad (7)$$

while the narrowing of the small polaron band at $E_{P0}$ is given by

$$\Delta E_{P0} = E_{band0} \exp(-2E_{JT0}/h\nu). \qquad (8)$$

The coupling energy $E_{JT0} = G_0^2/2K$ is the one of phonon coupling to states of the same band by symmetry-retaining vibrations. Throughout, $K = M\omega^2$ is the "stiffness" of the coupled vibration assumed the same for intraband and interband coupling.

May the co-operative Jahn-Teller effect [15] leading to a JT polaron is the one of mixing two degenerate carrier bands #1 and #2 of the same parity and bandwidth: $E_{band1} = E_{band2}$. The coupling energy $E_{JT12} = G_{12}^2/2K$ is the mixing energy of the latter bands where $G_{12}$ is the mixing constant. Following the above lines of reasoning, we set substituting for $E_{band0}$: $E_{band12} = E_{band1} + E_{band2}$ and, therefore, for a Jahn-Teller polaron the subsequent quantities should be

$$E_{P12} = -\tfrac{1}{2} E_{band12} - E_{JT12} + \tfrac{1}{2} h\nu, \qquad (9)$$

and also

$$\Delta E_{P12} = E_{band12} \exp(-2E_{JT12}/h\nu). \qquad (10)$$

Now, if the respective coupling constants and energies are of the same order of magnitude whether intraband or interband, then we get $\Delta E_{P12} \sim 2\Delta E_{P0}$. We arrive at the important conclusion that an itinerant JT polaron may be more mobile than Holstein's because of its wider polaron band. A related presumption had at the time led to the experimental discovery of a high-$T_c$ superconductivity in $La_{2-x}Sr_xCuO_4$.

## 3. Concluding discussion

Another set of parameters is listed in the caption to Figure 4. It has produced the fit of the theoretical formula to the experimental data which fit is rather acceptable. Moreover it suggests as a possibility quantal diffusion at low temperatures which has not been reported so far for AgBr. Even a quick glimpse at Figure 4 shows that the classical approach to the experimental diffusion data is less likely. The fit in Figure 4 has been obtained by four basic parameters: $E_B$, $h\nu$, $\eta$ and $\Lambda$ for an isothermic process ($Q = 0$), all the rest are derivatives.

An essential problem to arise as regards the carrier bands is the symmetry of the polaron or, else, whether we deal with Holstein- or JT- polaron: in other words, just how to distinguish between the two. A Holstein polaron arises from the coupling of a single electronic band to symmetry-retaining vibrations. On the contrary, the JT polaron arises from the mixing of two degenerate electron bands by symmetry-breaking vibrations [9]. An essential question is whether the outermost $7s^2$ electron of the $Ag^{++}$ ($Ag_{Ag}{}^\bullet$) ion can form equivalent states to give rise to two or more (e.g. six in an $O_h$ lattice) degenerate JT bands. Or, else, whether $Br_{Br}{}^\bullet$ can likewise regroup to form degenerate JT hole bands. But, because of the $O_h$ symmetry, the number of equivalent positions (alias degenerate states) is six with a hole circumventing over them around a central $Ag_{Ag}{}^x$ ion. We stress that the symmetry properties of an ion in a lattice depend essentially on the central symmetry of the surrounding medium.

We do not yet know which even parity vibration couples to the $6Br_{Br}{}^x$ octahedron around the central $Ag_{Ag}{}^\bullet$ ion ($E_g$ and $T_{2g}$ are two possibilities leading to distortions of the octahedron). In general, identifying the vibrational modes coupling to charge carriers in $O_h$ group materials is not that popular for siver halides as it is for high-$T_c$ cuprates. $T_{2g}$ or $E_g$ distortions would migrate across the crystal coupled to the $Ag_{Ag}{}^\bullet$ holes to form JT vibronic polarons. Similar conjectures may be applied relative to holes in the bromine band ($Br_{Br}{}^\bullet$). In any event, irrespective of the diffusing carrier nature, whether $Ag_{Ag}{}^\bullet$ or $Br_{Br}{}^\bullet$, the end result of bringing a hole to the activated surface will most likely be the same, since an $Ag_{Ag}{}^\bullet$ hole may recharge upon reaching the surface to give $Ag_{Ag}{}^\bullet + Br_{Br}{}^x \rightarrow Ag_{Ag}{}^x + Br_{Br}{}^\bullet$. The resulting bromine hole will undoubtedly bleach an activator silver atom outright.

We have presently outlined a statistical quantal rate theory for the diffusion of primary photocarriers in silver halides. Its most outstanding prediction is of a zero-point diffusion rate operative below 100 K. In any event, our calculations indicate a slow rise of diffusion constant as the temperature is advanced even below 100 K but the sensitivity was too poor to draw any definite conclusions. As a matter of fact, in a medium with favorable conditions for the propagation of Jahn-Teller distortions itinerant JT polarons will form and propagate. If the electronic band $E_{band}$ allows for an unimpeded electron motion, then the polaron coupling transforms the unit into a squeezed polaron band. The favorable condition will be met, provided the polaron band is not too narrow to allow for an unobstructed carrier migration.